\documentclass[prl,twocolumn,graphics]{revtex4}
\usepackage{graphicx}
\usepackage{color}

\begin{document}

\title{Exotic Superconducting Properties in the Electron-Hole Compensated Heavy Fermion `Semimetal' URu$_2$Si$_2$}
 
\author{Y.\ Kasahara$^\mathrm{1}$,  T.\ Iwasawa$^\mathrm{1}$, H.\ Shishido$^\mathrm{1}$, T.\ Shibauchi$^\mathrm{1}$,  K. Behnia$^{2}$,\\ Y.\ Haga$^\mathrm{3}$,\ T.\ D.\ Matsuda$^\mathrm{3}$, Y.\ Onuki$^\mathrm{3,4}$, M.\ Sigrist$^\mathrm{1,5}$, and Y.\ Matsuda$^\mathrm{1,6}$}
\affiliation{$^1$Department of Physics, Kyoto University, Kyoto 606-8502, Japan }%
\affiliation{$^2$Laboratoire de physique quantique (CNRS), ESPCI, 10 rue de Vauquelin, 75231 Paris, France}%
\affiliation{$^3$Advanced Science Research Center, Japan Atomic Energy Agency, Tokai 319-1195, Japan}%
\affiliation{$^4$Graduate School of Science, Osaka University, Toyonaka, Osaka 560-0043, Japan}%
\affiliation{$^5$Institute f\"ur Theoretische Physik, ETH Z\"urich, CH-8093, Switzerland}%
\affiliation{$^6$Institute for Solid State Physics, University of Tokyo, Kashiwanoha, Kashiwa, Chiba 277-8581, Japan}%


\begin{abstract}

We show that the charge and thermal transport measurements on ultraclean 
crystals of URu$_2$Si$_2$ 
reveal a number of unprecedented superconducting properties. The uniqueness is best highlighted by the peculiar field dependence of  thermal conductivity including the first order transition at $H_{c2}$ with a reduction of entropy flow.
This is a consequence of multi-band superconductivity with compensated electronic structure in the hidden order state of this system. 
We provide strong evidence for a new type of unconventional superconductivity with two distinct gaps having different nodal topology.

\end{abstract}

\pacs{71.27.+a,74.20.Rp,74.25.Fy,74.70.Tx}

\maketitle
The heavy-Fermion compound URu$_2$Si$_2$  has mystified researchers since the superconducting state ($T_c$~=~1.5~K) is embedded within the ``hidden order" phase ($T_h$~=~17.5~K) \cite{Palstra,Maple,Schlabitz}.
Although several exotic order parameters have been proposed for the  hidden order phase 
\cite{Shah}, 
it is not identified yet. 
According to several experimental observations, most of the carriers disappear below $T_h$ resulting in a density one order of magnitude smaller than in other heavy-Fermion superconductors \cite{Schoenes,Bel,Beh05}.  
Superconductivity with such a low density is remarkablesince the superfluid density is very low in some way reminiscent of underdoped cuprates. 
Moreover, pressure studies reveal that the superconductivity coexists with the hidden order but is suppressed by antiferromagnetic ordering \cite{Amitsuka}.

In this Letter, using ultraclean single crystals, we report various anomalous superconducting properties in  URu$_2$Si$_2$.  We show that a peculiar electronic structure appearing below the hidden order transition 
provides an intriguing stage on which a new type of unconventional superconducting state appears.

Single crystals of URu$_2$Si$_2$ were grown by the Czochralski pulling method in a tetra-arc furnace.
The well-defined superconducting transition was confirmed by the specific heat measurements. The thermal conductivity $\kappa$ was measured using a standard four-wire steady state method along the $a$-axis (heat current {\boldmath $q$}$\parallel a$). The contact resistance at low temperatures is less than 10~m$\Omega$.

We first discuss the electronic structure below $T_h$.  Figure 1 shows the temperature dependence of the resistivity $\rho$ along the $a$-axis and Hall coefficient $R_H$  (solid circles) defined as $R_H \equiv \frac{d\rho_{xy}}{dH}$ at $H \rightarrow 0$~T for {\boldmath $H$}$\parallel c$ in the tetragonal crystal structure.   In zero field, $\rho$ depends on $T$ as $\rho=\rho_0+AT^2$ below 6~K down to $T_c$.  The exceptionally low residual resistivity $\rho_0$~=~0.48$~\mu \Omega$~cm and large residual resistivity ratio $RRR$~=~670 attest the highest crystal quality currently achievable.  $R_H$ exhibits an eight-fold increase below $T_h$ and attains a $T$-independent value at low temperatures, associated with a strong reduction of the carrier density.   Most remarkably, the magnetoresistance (MR) increases with decreasing temperature and becomes extremely large at the lowest temperatures.    The inset of Fig.~1 shows the MR
$\Delta \rho/\rho_0\equiv (\rho(H)-\rho_0)/\rho_0$ extrapolated for $T\rightarrow 0$~K as a function of $H^2$.  The MR 
exhibits a nearly perfect $H^2$-dependence without 
saturation up to 10~T. At $\mu_0H$~=~10~T$, \Delta \rho/\rho_0$ reaches as large as 300. We note that the observed MR 
is much larger than in any known superconducting material.

The combination of the peculiar electronic structure of URu$_2$Si$_2$, i.e. high purity, the low carrier density and essentially equal number of electrons and holes (compensation), $n_e=n_h$, gives rise to this exceptionally large MR 
\cite{Pippard}. 
The observed quadratic field dependence of the MR 
without saturating behavior demonstrates the nearly perfect compensation.
is given by
\begin{equation}
\Delta \rho/\rho_0=(\omega_c^e\tau_e)\cdot (\omega_c^h\tau_h)
\end{equation}
and the Hall resistivity as
\begin{equation}
\rho_{xy}=\frac{H}{n_he}\cdot\frac{\omega_c^h\tau_h-\omega_c^e\tau_e}{\omega_c^h\tau_h+\omega_c^e\tau_e},
\end{equation}
where $\omega_c=eH/m$ is the cyclotron frequency for carriers of the mass $m$ and the scattering time $\tau$ \cite{Pippard}.  The suffixes $e$ and $h$ denote `electron' and `hole', respectively.  From Eq.~(1), we estimate $(\omega_c^e\tau_e) \cdot (\omega_c^h\tau_h) \simeq$~25 at $H_{c2}^c=$~2.8~T, where $H_{c2}^c$ is the upper critical field for {\boldmath $H$}$\parallel c$.  We note that this large $\omega_c \tau$-value sets a record of type-II superconductors.  The best-known systems exhibiting a similar behavior are very pure semimetals, such as bismuth and graphite \cite{Du05}.  The 
$T$-independent positive $R_H$ at low temperatures together with nearly $H$-linear $\rho_{xy}$ up to 10~T (not shown) indicates that $\omega_c^h \tau_h \gg  \omega_c^e \tau_e$,  or $\tau_e$ and $\tau_h$ have the same $T$-dependence. In either case, $1/R_He$ gives an upper limit of the hole number. The number of holes estimated from $R_H=2.3\times10^{-8}~$m$^3$/C at low temperatures, where the influence of the skew scattering can be neglected,  is $n_h\alt 2.6\times10^{26}$~m$^{-3}$ (corresponding to 0.021 holes per U-atom). Such a low carrier density implies Fermi surfaces occupying only a small part of the Brillouin zone.  According to de Haas-van Alphen (dHvA) experiments \cite{Ohkuni,Ber97},  one hole band with nearly spherical shape is reported.  This band with Fermi momentum $k_F\simeq1.8\times10^{9}$~m$^{-1}$ contains $\sim$~0.017 holes/U. This number gives a lower limit of hole number, because of the presence of very small hole pockets.  Thus we obtain $0.017 \alt n_h \alt 0.021$ holes/U.

\begin{figure}[t]
\includegraphics[width=9cm]{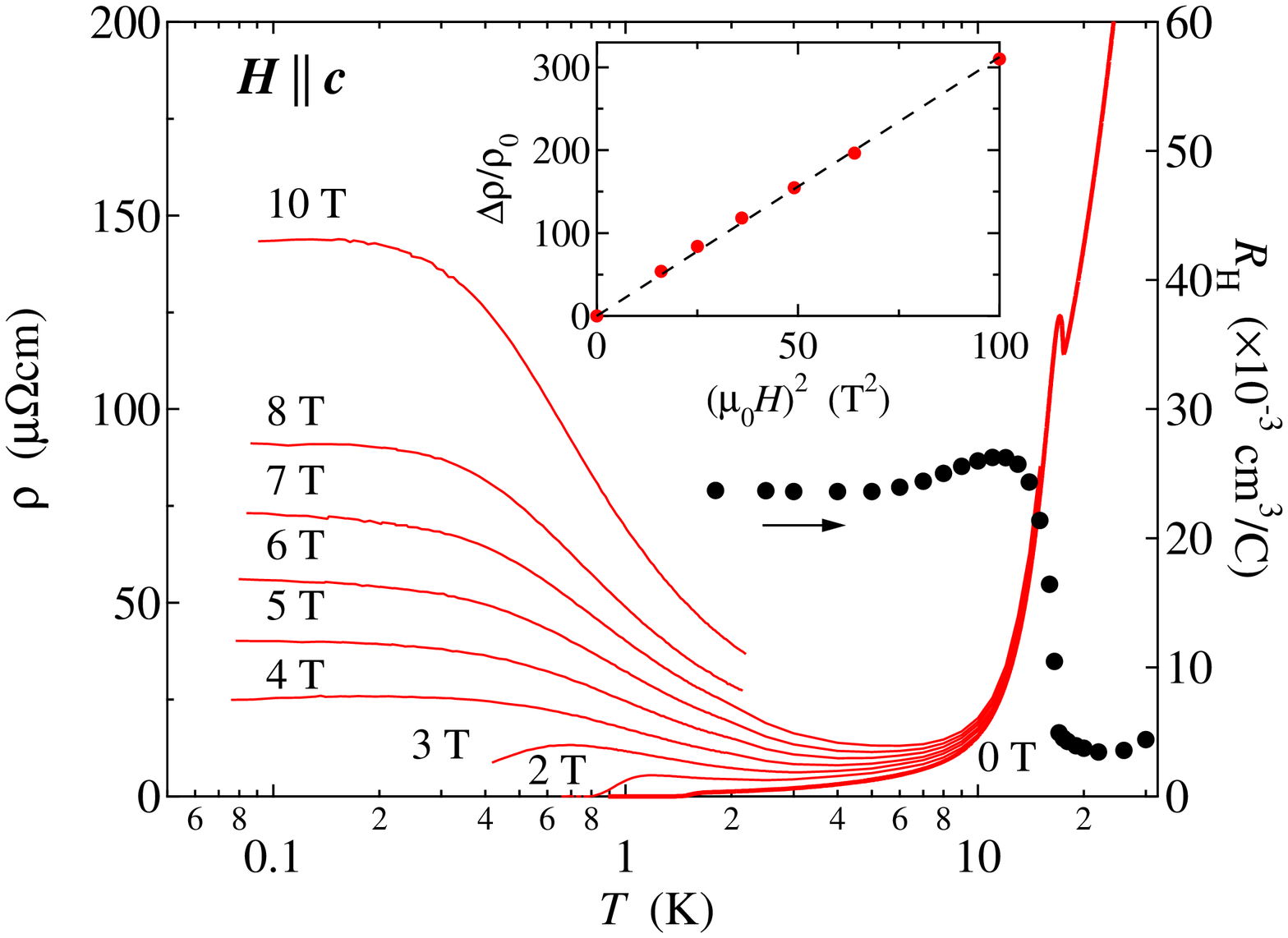}
\caption{(color online). Main panel: Temperature dependence of  $\rho$ (solid lines) along the $a$-axis in zero and finite magnetic fields ({\boldmath $H$}$\parallel c$) and  Hall coefficient $R_H$ (solid circles).  Inset:  The MR 
at $T \rightarrow 0$~K as a function of $H^2$.}
\end{figure}

The compensation necessarily indicates the presence of electron band, although it is not confirmed by the dHvA experiments \cite{Ohkuni,Ber97}. The magnitude of the electronic specific heat \cite{Maple} is five times larger than the estimation taking only the hole band assuming $m_h=13m_0$ \cite{Ohkuni}. This indicates that the electron mass is much heavier than hole mass. Moreover, the observed large anisotropies of the MR 
and the upper critical fields \cite{Ohkuni} indicates that the shape of the heavy electron band is elliptical \cite{Fermi}. We note that heavy electrons are consistent with the Hall effect which is dominated by light holes.

We next address the superconducting state.  The inset of Fig.~2 shows the $T$-dependence of the thermal conductivity divided by $T$, $\kappa/T$.  The Wiedemann-Franz (WF) ratio $L=(\kappa/T)\rho$ at $T_c$ is $(1.00\pm0.02) L_0$, where $L_0=2.44\times 10^{-8}~\Omega$W/K is the Sommerfeld value, indicating that the electron contribution dominates over the phonon contribution.  Below $T_c$,  $\kappa/T$ displays a steep increase and exhibits a pronounced maximum at 0.5~K, which arises from the strong suppression of the quasiparticle (QP) scattering rate due to the formation of a gap, as observed in several strongly correlated systems \cite{Kasa06}. 

Figure 2 depicts $\kappa/T$ plotted as a function of $T^2$ at very low temperature.  In zero field, $\kappa/T$ is well fitted by $\kappa/T=\kappa_{00}/T+bT^2$, where $b$ is a constant.  There is a fundamental difference in the low-temperature heat transport  between superconductors with  isotropic and anisotropic gaps.  In the latter, the QP heat conduction is entirely governed by nodal topology \cite{Mat05}.    As shown in Fig.~2, the presence of a residual value in $\kappa/T$ at $T\rightarrow 0$~K, $\kappa_{00}/T$,  is clearly resolved in zero field.  This indicates the presence of the residual normal fluid, which can be attributed to the presence of nodes in the gap function.  A more detailed analysis will be given later.

\begin{figure}[t]
\includegraphics[width=8cm]{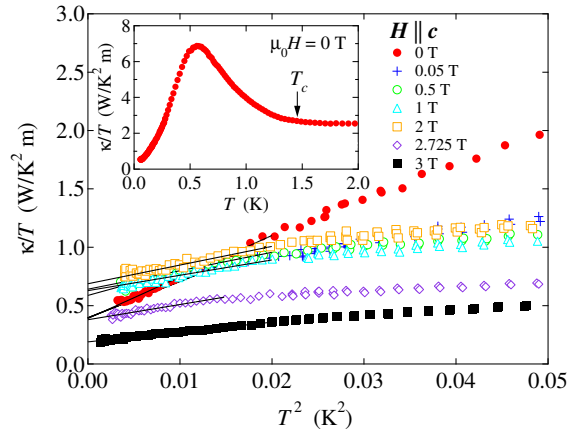}
\caption{(color online). Main panel: $\kappa/T$ plotted as a function of $T^2$ at very low temperatures. The data in magnetic fields ({\boldmath $H$}$\parallel c$) are obtained under field cooling conditions.   The lines are guides for eyes. Inset: 
$\kappa/T$ vs $T$ at high temperatures.} 
\end{figure}

The $H$-dependence of $\kappa(H)/T$ obtained by the extrapolation to $T\rightarrow 0$~K 
at each field value (see Fig.~2) is quite unusual, as shown in Figs.~3 (a) and (b).  There are three characteristic fields denoted as $H_{s}\simeq$~0.4~T, $H_{c2}^c\simeq$~2.8~T and $H_{c2}^{a}\simeq$~12~T, where $H_{c2}^{a}$ are the upper critical field for {\boldmath $H$}$\parallel a$.  For both {\boldmath $H$}$\parallel a$ and {\boldmath $H$}$\parallel c$, $\kappa(H)/T$ exhibits a pronounced increase below $\sim H_s$ (Fig.~3(b)), followed by a plateau-like behavior above $\sim H_s$.  Interestingly,  in spite of the large electronic anisotropy,  $\kappa(H)/T$ is nearly isotropic with respect to the field direction at low fields, as shown in Fig.~3(b).    For {\boldmath $H$}$\parallel c$,  $\kappa(H)/T$  exhibits a jump to a very low value, which is nearly 1/4 of $\kappa(H)/T$ just below $H_{c2}^c$, in a narrow field window $(\Delta H<0.2$~T) at $H_{c2}^c$.    The dashed blue and dotted red lines represent the expected $\kappa(H)/T$ by the WF-law for {\boldmath $H$}$\parallel c$ and {\boldmath $H$}$\parallel a$, respectively, assuming the quadratic dependent MR
(see the inset of Fig.~1).  Above $H_{c2}^c$, $\kappa(H)/T$ decreases rapidly with $H$ related to the remarkable MR, 
following the WF law $\kappa(H)/T=L_0/\rho(H)$.    On the other hand, for {\boldmath $H$} $\parallel a$, $\kappa(H)/T$  increases steeply with $H$ with downward curvature over a wide range of fields approaching $H_{c2}^{a}$.   Near $H_{c2}^{a}$, $\kappa(H)/T$ turns to decrease.   It is natural to assume that above our experimental field limit (12.4~T), $\kappa(H)/T$ drops to a lower value obtained by the WF law, similar to the case for {\boldmath $H$}$\parallel c$. 

\begin{figure}[t]
\includegraphics[width=9cm]{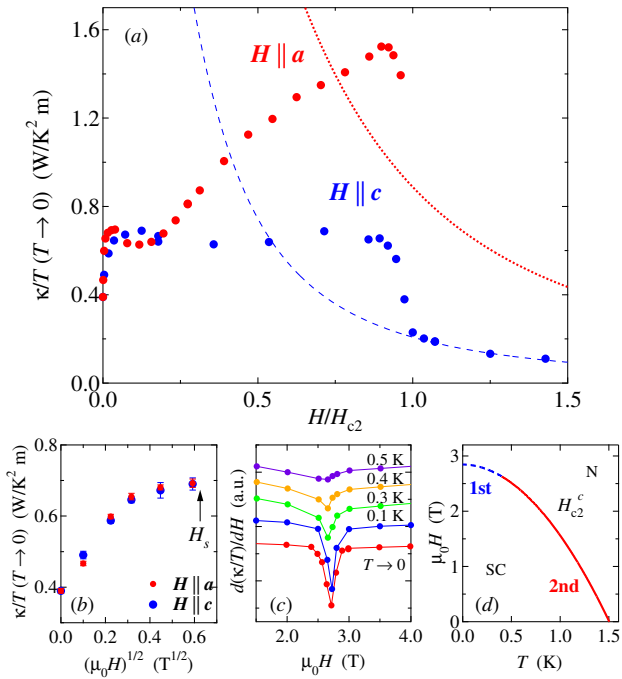}
\caption{(color online). (a) The field dependence of $\kappa/T$ for {\boldmath $H$}$\parallel c$ (blue circles) and {\boldmath $H$}$\parallel a$ (red squares) as a function of $H/H_{c2}^c$ and $H/H_{c2}^{a}$, respectively. 
The dashed blue and dotted red lines represent the expected $\kappa(H)/T$ by the WF-law for ({\boldmath $H$}$\parallel c$) and ({\boldmath $H$}$\parallel a$), respectively.  (b)  $\kappa(H)/T$ at low fields as a function of $\sqrt{H}$.   The characteristic field, $H_s=$0.4~T is indicated by the arrow. (c) The field derivative of $\kappa(H)/T$ for {\boldmath $H$}$\parallel c$  plotted as a function of $H$. (d) The $H-T$ phase diagram for {\boldmath $H$}$\parallel c$.   The transition between superconducting (SC) and normal (N) states is of first order below $\sim0.4$~K.}
\end{figure}

To our knowledge, the remarkable step-like reduction of $\kappa(H)/T$ at upper critical fields is unprecedented in type-II superconductors.   Since the thermal conductivity is related to an entropy flow, its jump (irrespective of its directions) immediately indicates the occurrence of the first order phase transition, as reported in CeCoIn$_5$ \cite{Izawa_CeCoIn5}.   To determine the end point of the first order transition,  we plot $d(\frac{\kappa}{T})/dH$  vs. $H$ at temperatures below 0.5~K for {\boldmath $H$}$\parallel c$ in Fig.~3(c).  A distinct dip structure  can be seen  below 0.4~K but is absent at 0.5~K, indicating that the end point is at $\sim 0.5$~K.  Figure 3(d) illustrates the $H-T$ phase diagram for {\boldmath $H$}$\parallel c$.   Since the Maki parameter of URu$_2$Si$_2$ is more than unity for both field directions, the origin of the first-order transition is most likely due to the Pauli paramagnetic limiting.  This suggests that URu$_2$Si$_2$ realizes an even-parity spin-singlet pairing state.

Usually, the heat conduction just below the upper critical field $H_{c2}$ is smaller than that in the normal state, owing to the reduction of number of QPs carrying heat and the enhancement of the scattering rate  by vortices. Therefore the observed field induced transition at $H_{c2}$ of URu$_2$Si$_2$ with an enhanced entropy flow in the superconducting state compared to the normal state is highly unusual. 
We point out that the compensation effect provides an explanation.  In the normal state, the compensation leads to an extremely large MR 
and hence large reduction of the thermal conductivity with increasing $H$.  However, in the superconducting mixed phase, the compensation of hole- and electron-like QPs is violated because the size and the nodal structure of the gaps on the two Fermi surfaces are very different. This implies that the heat transport is drastically changed at $H_{c2}$. 

The initial steep increase and subsequent plateau behavior of the thermal conductivity shown in Figs.~3(a) and (b) indicate that a substantial portion of QPs is already restored at 
$H_s$ much below the upper critical fields. 
Such a field dependence bears a strong resemblance to those of MgB$_2$ \cite{Sologubenko} and PrOs$_4$Sb$_{12}$ \cite{Seyfarth}, providing evidence for multi-band superconductivity in URu$_2$Si$_2$; $H_{s}$ is interpreted as a virtual upper critical field  that controls the field dependence of the smaller gap.  The multi-band superconductivity is, of course, compatible with the compensation effect originating from the presence of two bands.  It is natural to assume that the smaller gap originates from the light hole band.  In addition, the spherical structure of hole band is consistent with nearly isotropic $\kappa(H)/T$ below $H_s$.

The steep increase of $\kappa(H)/T$ at low fields is in strong contrast to the behavior of fully gapped superconductors, where all the QP states are bound to vortex cores and, therefore, the applied magnetic field hardly affects the thermal conduction except in the vicinity of $H_{c2}$. By contrast, in superconductors with nodes, the heat transport is dominated by the delocalized QPs and the Doppler shift of the QP spectrum induces a remarkable field dependence of $\kappa(H)/T$.  For line nodes, the QP density of states $N(0)$ increases with $\sqrt{H}$ \cite{Mat05}. As shown in Fig.~3(b), $\kappa(H)/T \propto N(0)$ for both field directions is nearly proportional to $\sqrt{H}$ up to $\sim H_s$. The presence of line nodes is further supported by a finite $\kappa_{00}/T$ term in zero field.   For line nodes, $\kappa_{00}/T$ is  given by $\frac{\kappa_{00}}{T}\simeq \frac{L_0}{\rho_0}\cdot\frac{\xi_{ab}}{\ell_e}$, where $\xi_{ab}$ and $\ell_e$ are the coherence length and mean free path within the $ab$-plane, respectively \cite{Tai97}.  To estimate $\xi_{ab}$, we use $H_s$ instead of $H_{c2}^c$, assuming the line nodes in the smaller gap.   Using $\ell_e\sim1.2~\mu$m in our crystal \cite{mean}, $\kappa_{00}/T$ is estimated to be 0.15~W/K$^2$m.  This value is in the same order with the observed $\kappa_{00}/T=$~0.38~W/K$^2$m.  These results suggest the presence of line nodes in the gap of the light hole band.

We next discuss the gap structure of heavy electron band.  This can be inferred from the thermal conductivity well above $H_s$ where essentially all QPs of the light band (small gap) contribute to the heat transport.   The plateau-like behavior of $\kappa(H)/T$ up to $H_{c2}^c$  for {\boldmath $H$}$\parallel c$  indicates that the QPs of the heavier band do not experience the Doppler shift.   On the other hand,  for {\boldmath $H$}$\parallel a$, the steep increase of $\kappa(H)/T$ with downward curvature above $\sim 0.2H_{c2}^a$ up to $H_{c2}^{a}$  can be interpreted as a Doppler shift effect for the QPs due to nodes. Note that in fully gapped superconductors, $\kappa(H)/T$ increases steeply only just below the upper critical field with an upward curvature.  For complete understanding of the multi-band effects,  the quantitative description for the crossover behavior from plateau to Doppler regimes for {\boldmath $H$}$\parallel a$ is required.  However, the fact that the Doppler shift occurs only in parallel field suggests that it originates from point nodes in the gap along the $c$-axis as these would not yield a Doppler shift for 
fields along the $c$-axis. 
Line nodes in contrast would give rise to a Doppler shift for any field direction \cite{Mat05}. 
In this way 
we distinguish the node topology in the gaps on the two bands.

\begin{figure}[t]
\includegraphics[width=7cm]{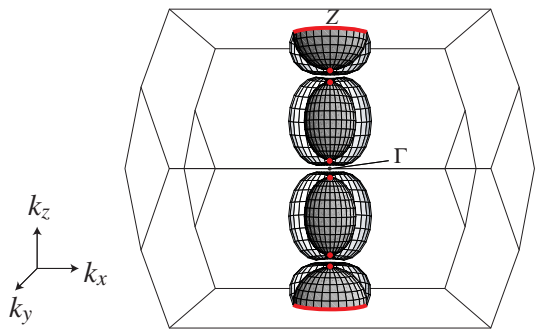}
\caption{(color online). The schematic figure of the Fermi surface (opaque) and superconducting gap structure (transparent) inferred from the present experiments \cite{FS}. 
}
\end{figure}

The present experiments strongly suggest (1) line node in the spherical light hole band, (2) point nodes in the elliptical heavy electron band (see Fig.~3(a)), and (3) spin-singlet pairing.  Although the detailed Fermi surface topology has not been fully determined, it is tempting to speculate on the possible gap symmetry which is consistent with the experimental observations. The classification scheme of even parity Cooper pairs allows for one type of state which has simultaneously (symmetry protected) line and point nodes. This is a two-component order parameter combined in a time reversal symmetry violating combination of the basic $d$-wave form $ \Delta_{k} \propto k_z (k_x \pm i k_y ) $. As the nodal structure is different on the two types of Fermi surfaces, it is necessary to account for the behavior of gap in the entire Brillouin zone. For the body-centered tetragonal crystal lattice the basic form of the gap function with the correct periodicity in $k$-space is given by
\begin{equation}
\Delta=\Delta_0\sin\frac{k_z}{2}c\left(\sin \frac{k_x+k_y}{2}a+i\sin \frac{k_x-k_y}{2}a\right),
\end{equation}
resulting from (nearest neighbor) interlayer pairing.  A schematic form of the gap structure and the Fermi pockets in the Brillouin zone is depicted in Fig.~4 \cite{FS}. The line nodes lie on the zone boundary on the hole-like light band, while both Fermi surfaces have point nodes along the $c$-axis. In avoiding a Fermi surface centered at the $ \Gamma $-point the line node for $ k_z = 0 $ does not play a role. This superconducting phase 
possesses a two-component order parameter which may give rise to interesting further features beyond the nodal properties of the QP gap.

In summary, the charge and thermal transport measurements on ultraclean URu$_2$Si$_2$  reveal a number of unprecedented superconducting properties.
We provide evidence for  superconductivity with two distinct gaps having different nodal topology. This intriguing superconducting state adds a unique 
example to the list of unconventional superconductors.

 We thank  H.~Amitsuka, S.~Fujimoto, H.~Ikeda, H.~Kontani, K.~Machida, K.~Maki, T.~Sakakibara, P.~Thalmeier and I.~Vekhter for discussion.
Recently we noticed the specific heat study which also suggests the point nodes \cite{Yano}.
%


\begin{thebibliography}{99}
\bibitem{mean}$\ell_e$ is obtained from Ref.\cite{Ohkuni}, taking into account the ratio of $\rho_0$ values in different crystals.
\bibitem{FS}
The dHvA experiments show the presence of the other two very small electron bands \cite{Ohkuni,Ber97}, which do not contribute to the transport and thermodynamic quantities.
\end{thebibliography}
\end{document}